\documentclass[aps,prl,reprint,superscriptaddress]{revtex4-2} %,nofootinbib
\pdfoutput=1

\usepackage{epsfig} 
\usepackage{amstext,amsmath,amssymb,amsfonts,amsmath,amsthm}
\usepackage{xcolor} 
\usepackage[colorlinks, linkcolor=blue, citecolor=blue, urlcolor=blue]{hyperref}
\usepackage{graphicx} 
\usepackage{braket} 
\usepackage{dsfont} % for double stroke font, as in \mathds
\usepackage{color}

\newcommand{\be}{\begin{equation}}
\newcommand{\ee}{\end{equation}} 
\newcommand{\nn}{\nonumber}
\newcommand{\f}{\frac}
\newcommand{\p}{\partial}
\newcommand{\la}{\langle}
\newcommand{\ra}{\rangle}

\newcommand{\rmd}{{\rm d}}
\newcommand{\im}{\mathrm{i}}

\DeclareMathOperator{\tr}{tr}
\newcommand{\hsp}{\hsigma_{+}}
\newcommand{\hsm}{\hsigma_{-}}

\renewcommand{\d}{\delta}

\renewcommand{\k}{\kappa} 
\renewcommand{\l}{\lambda}

\newcommand{\s}{\sigma} 
  
\newcommand{\vph}{\varphi}
 
\newcommand{\D}{\Delta}

\newcommand{\cG}{\mathcal{G}}
\newcommand{\cH}{\mathcal{H}}

\newcommand{\cJ}{\mathcal{J}}

\newcommand{\cO}{\mathcal{O}}

\newcommand{\cV}{\mathcal{V}}

\newcommand{\hsigma}{\hat{\sigma}}

\begin{document}

\title{\bf 1d Ising model with $1/r^{1.99}$ interaction}

\author{Dario Benedetti}
\affiliation{CPHT, CNRS, \'Ecole polytechnique, Institut Polytechnique de Paris, 91120 Palaiseau, France}
\author{Edoardo Lauria}
\affiliation{Laboratoire de Physique de l’École Normale Supérieure, Mines Paris, Inria, CNRS, ENS-PSL, Sorbonne Université, PSL Research University, Paris, France}
\author{Dalimil Maz\'{a}\v{c}}
\affiliation{
Institut de Physique Th\'{e}orique, Universit\'{e} Paris-Saclay, CEA, CNRS, F-91191 Gif-sur-Yvette, France}
\author{Philine van Vliet}
\affiliation{Laboratoire de Physique Th\'eorique de l'\'Ecole Normale Sup\'erieure, PSL University,\\CNRS, Sorbonne Universit\'es, UPMC Univ. Paris 06, 75231 Paris Cedex 05, France}

%\date{}

\begin{abstract}
We study the 1d Ising model with long-range interactions decaying as $1/r^{1+s}$. The critical model corresponds to a family of 1d conformal field theories (CFTs) whose data depend nontrivially on $s$ in the range $1/2\leq s\leq 1$. The model is known to be described by a generalized free field with quartic interaction, which is weakly coupled near $s=1/2$ but strongly coupled near the short-range crossover at $s=1$. We propose a dual description which becomes weakly coupled at $s=1$. At $s= 1$, our model becomes an exactly solvable conformal boundary condition for the 2d free scalar. We perform a number of consistency checks of our proposal and calculate the perturbative CFT data around $s=1$ analytically using both 1) our proposed field theory and 2) the analytic conformal bootstrap. Our results show complete agreement between the two methods.
\end{abstract}

\maketitle

%%%%%%%%%%%%%%%%%%%%%%%%%%%%%%%%%%%%%%%%%%%%%%%%%%%%%%%%

\paragraph{\bf Introduction.}
Statistical models with long-range interactions are of great theoretical interest due to their unusual features. They exhibit spontaneous symmetry breaking even in one dimension~\cite{Dyson:1968up}, yielding universality classes with continuously varying critical exponents~\cite{Fisher:1972zz} and exhibiting interesting crossover phenomena~\cite{Sak:1973,Brezin:2014,Behan:2017dwr,Behan:2017emf}. Long-range models also enjoy a fruitful interplay with boundary and defect conformal field theories~\cite{Paulos:2015jfa,Behan:2023ile} and even admit feasible experimental realizations~\cite{Defenu:2021glw}.

In this letter, we consider the 1d long-range Ising (LRI) model. On the lattice, it is described by the classical Hamiltonian:
\be
\cH_{{}_{\rm LRI}} = \f{ \cJ}{2} \sum_{i\neq j} \f{(\s_i-\s_j)^2}{|i-j|^{1+s}}\,,\quad \cJ>0\,,
\label{eq:Ising}
\ee
where $\s_i=\pm 1$ are the Ising variables at sites $i\in\mathbb{Z}$.
The parameter $s$ controls the range of the interaction. The model has a phase transition for $0<s\leq 1$~\cite{Dyson:1968up,Frohlich:1982,AizenmanChayes:1988}, with mean field critical exponents if $0<s\leq 1/2$ \cite{Aizenman:1988}, and no phase transition for $s>1$ \cite{Dyson:1969nonExist}; for $s=1$ the correlation length diverges at the transition, but the magnetization is discontinuous \cite{Thouless:1969,AizenmanChayes:1988}.

At the critical temperature, the long-distance (IR) dynamics of the 1d LRI is captured by a family of nonlocal 1d CFTs parametrized by $s\in(0,1)$, see~\cite{Paulos:2015jfa}. Conformal invariance is the statement that correlation functions transform covariantly under the group $\mathrm{PSL}_2(\mathbb{R})$ of fractional-linear transformations of the line. Because of reflection-positivity, local operators organize into unitary lowest-weight representations of $\widetilde{\mathrm{SL}}_2(\mathbb{R})$, labeled by the scaling dimensions $\Delta\geq 0$. The spin variables $\s_i$ give rise to a conformal primary operator $\sigma(x)$. Besides the conformal symmetry, the CFTs also possess a $\mathbb{Z}_2$ global symmetry, acting as $\sigma(x)\mapsto-\sigma(x)$, as well as parity symmetry $\sigma(x)\mapsto\sigma(-x)$.
Given the 1d setting and its few ingredients, this is probably the simplest family of interacting CFTs that has not been solved analytically.

The same family of 1d CFTs also arises as the IR fixed point of a generalized free field (GFF) $\varphi$ -- identified with $\sigma$ in the IR -- perturbed by a quartic interaction \footnote{Here $c_s = \Gamma(s+1)\sin(\frac{\pi s}{2})/\pi$ so that $c_s/|x|^{1+s}$ becomes $|p|^{s}$ in momentum space. 
}:
\be \label{eq:action-LRI}
\begin{split}
S_{{}_{\rm LRI}}[\vph] =& \f{c_s}{4}  \int_{-\infty}^{+\infty} \rmd x_1 \rmd x_2 \f{(\vph(x_1)-\vph(x_2))^2}{|x_1-x_2|^{1+s}} \\
& + \int_{-\infty}^{+\infty} \rmd x \left(  \f{\l_2}{2} \vph(x)^2 + \f{\l_4}{4} \vph(x)^4 \right) \;.
\end{split}
\ee
The non-local kinetic term is not renormalized, and thus the scaling dimension sticks to its UV value, i.e.\ $\Delta_\sigma=\Delta_{\varphi}=\frac{1-s}{2}$~\cite{Lohmann:2017}. For $0<s\leq 1/2$, the quartic interaction is irrelevant and the IR CFT is thus equivalent to the GFF. For $1/2< s <1$, $\varphi^4$ is relevant and the IR CFT is interacting~\cite{Fisher:1972zz}. The non-local equation of motion $\partial^s\varphi\sim\varphi^3$ then predicts the presence of a $\mathbb{Z}_2$-odd parity-even primary operator $\chi\sim\varphi^3$ of protected dimension $\Delta_{\chi}=\frac{1+s}{2}$~\cite{Paulos:2015jfa,Behan:2017dwr,Behan:2017emf}. In fact, $\sigma$ and $\chi$ are related by the shadow transform, leading to nonperturbative relations between different OPE coefficients~\cite{Paulos:2015jfa,Behan:2018hfx}, see Appendix~A for details.

For $s-1/2\ll 1$, the description~\eqref{eq:action-LRI} is weakly coupled at the critical point, but becomes strongly coupled as $s$ increases towards $1$. Therefore, studying the properties of the 1d LRI CFT near $s=1$ is a hard problem.

We can gain some intuition into the physics near $s=1$ from the behaviour of the LRI model in dimensions $d>1$. As explained in~\cite{Behan:2017dwr,Behan:2017emf}, the theory with $1/r^{d+s}$ interactions has a dual description as the short-range Ising (SRI) coupled to the GFF $\chi$ via $\sigma\chi$. The latter becomes weakly coupled near $s^\star(d) = d-2\Delta_\sigma^\star(d)$, where $\Delta_\sigma^\star(d)$ is the dimension of the spin field in the critical SRI model \footnote{The LRI model has also been studied up to three loops near $s=d/2$~\cite{Benedetti:2020rrq,Benedetti:2024mqx,Behan:2023ile,Rong:2024vxo}.}.

The proposal~\cite{Behan:2017dwr,Behan:2017emf} breaks down in $d=1$ because the SRI model does not correspond to a CFT. Instead, it exhibits a first-order phase transition at zero temperature, described by a TQFT with a pair of ground states $|\pm 1\rangle$. The spin field is a topological operator acting as $\sigma |a \rangle = a |a\rangle$. In particular, $\Delta^\star_\sigma(1) = 0$, which together with $\Delta_{\sigma}=\frac{1-s}{2}$ correctly reproduces the crossover location $s^\star(1)=1$. By analogy with $d>1$, one may guess a description near $s=1$ in terms of the GFF $\chi$ and the topological operator $\sigma$ coupled via $\sigma\chi$. We will see below that the correct answer is similar but considerably richer.

To solve the IR dynamics of the 1d model near $s=1$, we need to identify the weakly coupled degrees of freedom. In terms of~\eqref{eq:Ising}, these are the locations of the domain walls, i.e.\ the sites $i$ where $\sigma_i$ flip from $-1$ to $1$, or vice versa. Anderson and Yuval~\cite{Anderson:1971jpc} observed that domain walls are dilute at low temperature and rewrote the $s=1$ LRI model as a Coulomb gas of alternating kinks and antikinks. Later, Kosterlitz~\cite{Kosterlitz:1976zz} extended their analysis to small positive values of $1-s$.

While the Anderson-Yuval-Kosterlitz (AYK) model correctly captures the physics, it is rather cumbersome to work with. As a result, it has not been used to systematically solve the critical LRI model near $s=1$. In this letter, we remove this obstacle by recasting the AYK model as a bona fide quantum field theory.

Indeed, we will argue that the critical 1d LRI model for $s\leq 1$ is equivalent to the IR fixed point of a GFF of negative dimension $-\frac{1-s}{2}$ and compact target space, coupled to a single qubit. The qubit is a remnant of the 1d Ising TQFT. The theory is reflection-positive thanks to compactness of the GFF theory.

The fixed-point couplings vanish as $s\rightarrow 1$, where our model becomes equivalent to an exactly solvable conformal boundary condition for the 2d free compact scalar. In particular, we find that all scaling dimensions of the LRI model tend to integer values in this limit.

Our model makes a wealth of predictions for the perturbative CFT data near $s=1$, extending the results of Kosterlitz~\cite{Kosterlitz:1976zz}.

Finally, starting from the exactly solvable $s=1$ theory as a seed, we determine the perturbative CFT data for $s<1$ analytically from the conformal bootstrap using analytic functionals~\cite{Mazac:2016qev}. We find perfect agreement with RG computations in our model.

More details and further results will be presented in a forthcoming publication \cite{long}.

\paragraph{\bf Physics near the cross-over.}
The AYK model describes the physics of the phase transition in terms of domain walls between positive and negative spins. As emphasized in~\cite{Kosterlitz_2016}, this is analogous to the effective vortex theory for the 2d XY model. The configuration space of the AYK model consists of positions $x$ of kinks and anti-kinks, which must alternate. Introducing the UV cutoff $a$, the hamiltonian~\eqref{eq:Ising} gives rise to an attractive potential $ \cV_s(x) = (|x|^{1-s}-a^{1-s})/(1-s)$ between a kink and an antikink at distance $x$ and a repulsive potential $-\cV_s(x)$ between kinks of like charges. Note that $\cV_1(x)=\log(|x|/a)$. The resulting partition function is a sum over the number $n$ of kink-antikink pairs and an integral over their locations $x_1>x_2>\ldots>x_{2n}$, satisfying $x_i-x_{i-1}>a$:
\be \label{eq:Z_Kosterlitz}
Z^{\text{AYK}}_{s}(b,g) = \sum_{n=0}^{\infty} g^{2n}\!\!\! \int \prod_{i=1}^{2n} \!\rmd x_i
 \,e^{2 b^2 \sum\limits_{i<j} (-1)^{i-j} \cV_s\left(x_{ij}\right) }\,.
\ee
Here and in the following $x_{ij} = x_i-x_j$. The fugacity $g$ and the parameter $b^2$ are determined by $\cJ$ and a possible short-range term in the Hamiltonian~\eqref{eq:Ising}. Applying a basic version of Wilson renormalization, Kosterlitz~\cite{Kosterlitz:1976zz} found a fixed point at $g=O(\sqrt{\delta})$, $b^2 = 1+O(\delta)$, where we introduced $\delta = 1-s$. He further predicted the presence of two $\mathbb{Z}_2$-even operators $\cO_{\pm}$ of dimensions $\Delta_{\pm}=1\pm\sqrt{2\delta}+O(\delta)$ at the fixed point. Assuming no level-crossing, we can identify these with the lightest $\mathbb{Z}_2$-even operators of the $\varphi^4$ description: $\cO_-\sim\varphi^2$, $\cO_+\sim\varphi^4$. We will confirm and significantly extend these predictions.

\paragraph{\bf Our model.}
Consider the 1d GFF $\phi$ of negative dimension $\D_\phi=-\delta/2$ and a compact target space of radius $1/b_0$, i.e.~$\phi\sim\phi+2\pi n/b_0$ for $n\in\mathbb{Z}$. For $0<\delta<1$, this theory can be rigorously defined~\cite{Lodhia:2016fractional} as a probability measure on the space of continuous functions $\mathbb{R}\rightarrow S^1_{1/b_0}$.
Due to the identification in field space, all well-defined observables must be built out of $\partial^n_x\phi(x)$ with $n\geq 1$, and vertex operators $V_{n}(x)\equiv e^{\im n b_0 \phi(x)}$, with $n\in\mathbb{Z}$ \footnote{In \cite{Lodhia:2016fractional}, the probability measure is defined on functions $\mathbb{R}\rightarrow \mathbb{R}$ defined modulo an additive constant. Our theory is defined by making the constant mode compact.}.
Since $\phi(x)$ without derivatives is not a well-defined observable, the theory is reflection positive~\cite{Jorgensen:2018}. We will denote the expectation value in this GFF theory by $\langle\cdot \rangle_0$, so in particular $\la(\phi(x)-\phi(y))^2\ra_0\propto |x-y|^{1-s}$.

The Boltzmann weight of a configuration of kinks and anti-kinks in~\eqref{eq:Z_Kosterlitz} now precisely agrees with the correlation function of vertex operators with alternating charges $\langle V_{1}(x_1)V_{-1}(x_2)\ldots V_{1}(x_{2n-1})V_{-1}(x_{2n})\rangle_0$. To force the charges to alternate, we form a product of the GFF theory with the algebra of $2\times 2$ matrices and supplement $V_{\pm 1}(x)$ with $\hsigma_\pm = \f12( \hsigma_1 \pm \im \hsigma_2)$. Here $\hsigma_{i=1,2,3}$ are the Pauli matrices.

This leads to our central observation: The critical 1d LRI model can be identified with the IR fixed point of the following 1d continuum model:
\be \label{eq:Z}
Z_s(b,g) = \Big{\la}\!\tr\mathrm{P}\mathrm{exp} \Big{\{}\int\limits \left[g\, \cO_g(x)  + h\, \cO_h(x)\right]\rmd x\Big{\}}  \Big{\ra}_{0}, \\
\ee
where $b=b_0- \sqrt{2}h$, and we introduced the operators
\be
\cO_g(x) \equiv\hsigma_+ V_{1}(x)+ \hsigma_- V_{-1}(x)\,,\;
\cO_h(x) \equiv \frac{\im \hsigma_3}{\sqrt{2}}\partial_x \phi(x).\label{eq:operators}
\ee
Note that $\tr\mathrm{P}\mathrm{exp}$ in~\eqref{eq:Z} is the trace of the path-ordered exponential of the $2\times 2$ matrix in curly brackets.

The first justification of our claim is that we reproduce the AYK partition function~\eqref{eq:Z_Kosterlitz} by expanding~\eqref{eq:Z} in $g$. Indeed, since $\hat{\sigma}_{+}^2=\hat{\sigma}_{-}^2=0$, only alternating chains of kink and antikink operators $\hsigma_+ V_{1}(x)$ and $\hsigma_- V_{-1}(x)$ survive. The effect of $\cO_h$ is to shift $b_0$ to $b_0-\sqrt{2}h$. In other words, $h$ is a redundant coupling that  can always be traded for a shift in $b_0$. However, it is convenient to keep $h$ explicit in order to renormalize the model, as we will explain.

The partition function \eqref{eq:Z} %$\Tr \la   \cD  \ra$ 
is similar to those encountered in impurity models \cite{Cuomo:2022xgw,Bianchi:2023gkk}, and the path ordering of $2\times 2$ matrices can be traded for a path integral using a complex bosonic spinor $z(x)=\{z_1(x),z_2(x)\}$ subject to $\bar{z}(x)z(x)=1$, similarly to \cite{Clark_1997,Cuomo:2022xgw,Bianchi:2023gkk}. 

Since the covariance of the GFF $\phi$ agrees with that of a free field in $s+1=2-\delta$ dimensions, our model can be formally interpreted as a one-dimensional defect for the $(s+1)$-dimensional free theory. At $s = 1$, the bulk is 2d and the defect is a boundary condition, with the vertex operators becoming conformal primaries of the Gaussian theory.
The resulting $s=1$ model then becomes the bosonized version of the Kondo model~\cite{Schotte:1970,Fendley:1995}, already related to the LRI model by Anderson and Yuval~\cite{Anderson:1971jpc}. However, already at $s=1$, our ensuing RG analysis and identification of the spectrum of operators are novel.

For $s=1-\d<1$, the model~\eqref{eq:Z} is a genuinely new theory. It might seem rather unorthodox, since $\Delta_{\phi} <0$ and since the vertex operators do not have a definite scaling dimension in the UV. Nevertheless, it is consistent:  the GFF has been rigorously constructed, and the interacting theory can be studied perturbatively, as we will discuss. 
Notice that similar types of non-polynomial, non-scaling operators do appear in other contexts, such as in Lagrangians of nonlinear sigma models (e.g.\ \cite{ZinnJustin:2002ru}), as well as in the form of monopole interactions in low-energy effective actions of (bosonized) QED${}_3$ \cite{Komargodski:2017dmc,Aharony:2024ctf}.
However, unlike typical phenomenological Lagrangians, our model is power-counting renormalizable, due to the negative dimension of the GFF.

In the rest of the letter, we will perform a number of additional checks which confirm~\eqref{eq:Z} as a correct UV description of the critical 1d LRI model.

\paragraph{\bf Symmetries.}
For $g=h=0$, the UV theory~\eqref{eq:Z} has $\mathrm{O}(2)$ global symmetry acting on the target space circle of $\phi$, and $\mathrm{PU}(2)$ global symmetry acting on the $2\times 2$ matrix degrees of freedom $A$ by conjugation. By turning on $\cO_g$ and $\cO_h$, $\mathrm{O}(2)\times\mathrm{PU}(2)$ is broken to the diagonal $\mathrm{O}(2)$, generated by $\mathrm{U}(1)$ rotations and $\mathbb{Z}_2$ reflection
\begin{align}
&\mathrm{U}(1): &&\phi(x)\mapsto\phi(x)+\alpha/b_0\,, &&A\mapsto e^{-\im\frac{\alpha}{2}\hsigma_3}Ae^{\im\frac{\alpha}{2}\hsigma_3},\label{eq:U1}\\
&\mathbb{Z}_2:&&\phi(x)\mapsto-\phi(x)\,,&&A\mapsto \hsigma_1 A\hsigma_1\,,
\label{eq:Z2}
\end{align}
where $\alpha\in\mathbb{R}/2\pi\mathbb{Z}$. The model~\eqref{eq:Z} is natural because $\cO_g$ and $\cO_h$ are the only relevant or marginal operators invariant under this $\mathrm{O}(2)$. The model also respects parity $x\mapsto -x$. In order for it to be compatible with the path ordering, and to preserve $\cO_g$ and $\cO_h$, it must act as
\be
\text{parity}:\quad\phi(x)\mapsto-\phi(-x)\,,\quad A\mapsto A^T\,.
\label{eq:parity}
\ee
In the next section, we will confirm the existence of an IR fixed point of~\eqref{eq:Z} at $b=1+O(\delta)$, $g=O(\sqrt{\delta})$. The fixed-point theory thus exhibits $\mathrm{O}(2)\rtimes\text{parity}$ symmetry.

We claim that the sector of the fixed-point theory with vanishing charge under the $\mathrm{U}(1)$ in~\eqref{eq:U1} is exactly equivalent to the critical 1d LRI CFT. In 1d, restricting to uncharged operators is the same as gauging of the $\mathrm{U}(1)$ symmetry. Thus, we claim that the 1d LRI CFT coincides with the fixed point of~\eqref{eq:Z} after $\mathrm{U}(1)$ gauging.

As a first consistency check, note that the symmetry of the zero-charge sector is precisely $\mathbb{Z}_2\times\text{parity}$, acting as~\eqref{eq:Z2},~\eqref{eq:parity}. This precisely agrees with the symmetry structure of the Ising model. Note that both $\mathbb{Z}_2$ and parity switch kink and antikink operators $\hsigma_+ V_1(x)\leftrightarrow\hsigma_- V_{-1}(x)$, as they should. $\mathrm{U}(1)$ symmetry has no counterpart in the Ising model, so it should be gauged \footnote{The $\mathrm{U}(1)$ gauging also ensures that the theory depends on $b_0$ and $h$ only through $b_0-\sqrt{2}h$. Indeed, correlation functions are invariant under $(b_0,h)\mapsto(b_0+\sqrt{2}t,h+t)$ if and only if each insertion is $\mathrm{U}(1)$-neutral.}.

Furthermore, we will soon see that the spectrum of operators predicted by the $\mathrm{U}(1)$ gauging picture perfectly reproduces the expected spectrum of operators of the critical LRI model.

\paragraph{\bf Renormalization.}
The renormalization of the model \eqref{eq:Z} near $\d=0$ can be performed in the framework of conformal perturbation theory (see e.g.\ \cite{Komargodski:2016auf}), with added subtleties due to the path ordering, as we will detail in \cite{long}. Since $\la\tr(\cO_g\cO_g\cO_h)\ra_0\neq 0$, it is crucial to include the term $h \,\cO_h$ in \eqref{eq:Z}. Its renormalization induces an effective flow for $b$, and the latter cannot be captured by wave function renormalization.
Therefore, we fix $b_0=\k^{\d/2}$ for some arbitrary mass scale $\k$, renormalize $h$, and redefine the physical coupling in cutoff units  as $b=(b_0 -\sqrt{2} h)a^{\d/2}$.
In a minimal subtraction scheme, we obtain the $\beta$-functions
\be
\beta_g =  (b^2-1) g+g^3+\ldots\,,\quad
\beta_{b^2} =-\delta b^2 + 4 b^2 g^2+\ldots
\label{eq:beta}
\ee
up to higher-order terms in $g$, $b^2-1$ and $\delta$ \footnote{If $g=O(\sqrt{\delta})$ and $b^2-1=O(\delta)$ as $\delta\rightarrow 0$ (as true at the fixed point), then~\eqref{eq:beta} contain all the terms up to order $\delta^{2}$ for $\beta_g$ and up to order $\delta^{3/2}$ for $\beta_{b^2}$.}. These $\beta$-functions agree with~\cite{Kosterlitz:1976zz} and improve it by the $g^3$ term.
The UV fixed point is located at $g=b=0$, so that the theory is asymptotically free and noncompact, and thus UV complete.
The IR fixed point is located at $g_\star=\sqrt{\d/4}+O(\delta^{3/2})$, $b^2_\star =1- \d/4+O(\delta^2)$.
The IR fixed-point theory is weakly coupled for $\d\ll1$, justifying the perturbative treatment. It becomes complex for $\d<0$, in accordance with the absence of a second-order phase transition for $s>1$.

By linearizing around the fixed point, we identify the scaling operators $\cO_{\pm}= (\cO_{h}\pm\cO_{g})/\sqrt{2} + O(\sqrt{\delta})$ and compute their scaling dimensions
\be
\D_{\pm} = 1\pm\sqrt{2\delta}+\delta/4+O(\delta^{3/2}) \;.\label{eq:dimOpm}
\ee
The $O(\sqrt{\delta})$ term agrees with~\cite{Kosterlitz:1976zz} and the $O(\delta)$ term is a new result. The appearance of the fixed point in the setting of our field theory \eqref{eq:Z} opens the door to the computation of many other quantities via the renormalization group \cite{long}.

\paragraph{\bf Operator spectrum.}
Our model provides an exact solution of the 1d LRI CFT in the limit $s\rightarrow 1$. Indeed, the fixed point then approaches $b=1$, $g=0$. Furthermore, we have $\Delta_\phi\rightarrow 0$. Setting $h=0$ without loss of generality, the target space for $\phi$ has radius $1/b_0 = 1$, which is precisely the self-dual radius for the 2d free scalar. Therefore, before gauging the $\mathrm{U}(1)$ symmetry, the 1d GFF part of our model is equivalent to the Neumann boundary condition for the 2d free scalar at the self-dual radius. Its operator spectrum consists of products of $\partial^n\phi$ with $n\geq 1$ and vertex operators $e^{i n\phi}$, the latter having scaling dimension $n^2$.

According to our proposal, the $\delta\rightarrow 0$ limit of the 1d LRI CFT is thus described by the product of this boundary condition with the algebra of $2\times 2$ matrices, restricted to the sector invariant under the $\mathrm{U}(1)$ symmetry~\eqref{eq:U1}. As in~\eqref{eq:Z}, correlation functions are computed by $\langle\cO\rangle = \frac{1}{2}\tr\langle\mathrm{P}(\cO)\rangle_0$, where $\mathrm{P}$ stands for path ordering and $\langle\cdot\rangle_0$ for the expectation value in the Neumann b.c.

We can now write down the $s\rightarrow 1$ limit of the complete spectrum of the 1d LRI CFT. A basis of operators with zero $\mathrm{U}(1)$ charge and definite scaling dimension, parity, and $\mathbb{Z}_2$ charge consists of $\mathds{1}\cO$, $\hat{\sigma}_3\cO$ and $(e^{\im\phi}\hat{\sigma}_+\pm e^{-\im\phi}\hat{\sigma}_-)\cO$, where $\cO$ is an arbitrary operator of the form $:\!(\partial\phi)^{n_1}(\partial^2\phi)^{n_2}\ldots (\partial^N\phi)^{n_N}\!:$ with $n_j\in\mathbb{Z}_{\geq 0}$. In particular, we see that all scaling dimensions tend to integers as $s\rightarrow 1$. Table~\ref{tab:spectrum} shows the low-lying spectrum of primary operators and their quantum numbers.

\begin{table}[h!]
\be\nn
\begin{array}{c || c | c | c | c}
\Delta & d_{++}(\Delta) & d_{+-}(\Delta) & d_{-+}(\Delta) & d_{--}(\Delta) \\\hline
 0 & 1 & 1 & 0 & 0 \\
 1 & 2 & 1 & 0 & 1 \\
 2 & 1 & 1 & 0 & 0 \\
 3 & 2 & 1 & 0 & 1 \\
 4 & 2 & 3 & 1 & 0 \\
% 5 & 4 & 2 & 0 & 2 \\
\end{array}
\ee
\caption{Low-lying spectrum of primaries in the $s\rightarrow 1$ limit of the 1d LRI CFT, as predicted by our model. $d_{ab}(\Delta)$ denotes the number of linearly independent primaries of dimension $\Delta$ with charge $a$ under parity and $b$ under the global $\mathbb{Z}_2$.}
\label{tab:spectrum}
\end{table}
Besides the identity, the only operator with $\Delta=0$ is $\hat{\sigma}_3$. It is parity-even and $\mathbb{Z}_2$-odd, and we are therefore forced to identify it with the $s\rightarrow 1$ limit of the spin field $\sigma(x)$. The two uncharged operators with $\Delta = 1$ are precisely $\cO_h = \im\hat{\sigma}_3\partial\phi/\sqrt{2}$ and $\cO_g = \hat{\sigma}_{+}e^{\im\phi}+\hat{\sigma}_{-}e^{-\im\phi}$, which generate the RG flow to the critical theory for $s<1$. As we saw in the previous section, their linear combinations $\cO_{\pm}$ remain primaries for $s<1$. Next, the unique $\Delta = 1$, parity-even, $\mathbb{Z}_2$-odd primary is $\im\partial\phi/\sqrt{2}$. We must therefore identify it with $\chi$! For $s<1$, $\phi$ receives no wavefunction renormalization and thus $\Delta_{\chi} = \Delta_\phi + 1 = \frac{1+s}{2}$ is protected, as required. %
Finally, the unique $\Delta = 1$, parity-odd, $\mathbb{Z}_2$-odd primary is $\hat{\sigma}_{+}e^{\im\phi}-\hat{\sigma}_{-}e^{-\im\phi}$. The presence of a primary with these quantum numbers is required by continuity in $s$; as $s\rightarrow 1$, the conformal multiplet of $\sigma$ becomes reducible $[\frac{1-s}{2}]\rightarrow [0]\oplus [1]$. In other words, $\hat{\sigma}_{+}e^{\im\phi}-\hat{\sigma}_{-}e^{-\im\phi}$ becomes the first descendant of $\sigma$ for $s<1$. Due to the path ordering, descendants of $\sigma$ should be defined using a defect covariant derivative $D_x$~\cite{Bianchi:2023gkk}. In our case, we find $D_x \s = 2g (\hat{\sigma}_{+}e^{\im\phi}-\hat{\sigma}_{-}e^{-\im\phi})$. Furthermore, the Schwinger-Dyson equations of our model identify $\hat{\sigma}_{+}e^{\im\phi}-\hat{\sigma}_{-}e^{-\im\phi}$ as the shadow operator of $\phi$. This shows that $\Delta_\sigma = \frac{1-s}{2}$ is protected and that $\sigma$ and $\chi$ form a shadow pair as required.
Table~\ref{tab:duality} summarizes the identification of the first five nontrivial operators with their counterparts in the dual quartic model~\eqref{eq:action-LRI}.

For another consistency check, note that due to the OPE relations~\eqref{eq:opeRelation}, any parity-odd, $\mathbb{Z}_2$-even primary appearing in the $\sigma\times\chi$ OPE must have a protected dimension whose value is an even integer for all $s$ \footnote{This statement is the $d=1$ version of the $d>1$ fact that odd-spin primary operators in the $\sigma\times\chi$ OPE have protected dimensions, see~\cite{Behan:2018hfx,Behan:2023ile}.}. In the mean field theory at $s=1/2$, the first such operator occurs at $\Delta = 4$ and has the schematic form $\varphi^3\partial^3\varphi$. The $s\rightarrow 1$ limit of this operator can be cleanly identified as the only nonzero entry of the $d_{-+}$ column of Table~\ref{tab:spectrum}.

We see that the spectrum of our model is in perfect agreement with what is expected from the critical LRI model.

\begin{table}[h!]
\be\nn
\begin{array}{c | c | c | c|c}
	\text{operator}& \Delta & (-1)^J &\mathbb{Z}_2 &\text{dual}\\\hline
	%\id &0& +1 & +1 \\
	\sigma=\hat{\sigma}_3 &\qquad\qquad\; \delta/2  \hfill \text{ exact}& +1 & -1 & \vph\\
	\chi=\im\partial\phi/\sqrt{2} &\qquad\quad\;\; 1-\delta/2 \hfill\text{exact} & +1 & -1 & \vph^3\\
	\cO_- &1-\sqrt{2\delta}+\delta/4+O(\delta^{\frac{3}{2}}) & +1 & +1 & \vph^2\\
	\cO_+ &1 +\sqrt{2\delta}+\delta/4+O(\delta^{\f32}) & +1 & +1 & \, \vph^4\\
	\hsp e^{\im\phi}-\hsm e^{-\im\phi} &\qquad\quad\;\;1+\delta/2 \hfill \text{exact}& -1 & -1 & \p\vph \,
\end{array}
\ee
\caption{Perturbative and exact results at $s=1-\d$, derived using the model~\eqref{eq:Z}. $(-1)^J$ and $\mathbb{Z}_2$ are the charges under space parity and global $\mathbb{Z}_2$. The last column shows the corresponding operator in the quartic model~\eqref{eq:action-LRI}.}
\label{tab:duality}
\end{table}

\paragraph{\bf The Conformal Bootstrap.}
Since the critical LRI model defines a 1d CFT for any $0<s<1$, it can be studied using the conformal bootstrap. While the bootstrap was originally applied to CFTs in $d\geq 2$, it proves very powerful also in $d=1$ (see~\cite{Bianchi:2023gkk,Ferrero:2023gnu, Dey:2024ilw} for some recent examples).

Besides the spectrum of scaling exponents $\Delta_i$ of primary operators $\cO_i$, any 1d CFT is characterized by the OPE coefficients $c_{ijk}\sim\langle\cO_i\cO_j\cO_k\rangle$. Thanks to OPE associativity, this data satisfies an infinite set of nonperturbative \emph{crossing equations}, taking the schematic form
\be
\sum\limits_{m} c_{ijm}c_{k\ell m} G_{\Delta_m}(z) = 
\sum\limits_{m} c_{jkm}c_{\ell i m} G_{\Delta_m}(1-z)\,.
\label{eq:crossing}
\ee
Here $G_{\Delta}(z)$ are the conformal blocks and the equations are valid for all $i,j,k,\ell$ and all $z\in(0,1)$, see Appendix~A for details.

As our model shows, the CFT data (=the spectrum and OPE coefficients) of the LRI model vary continuously with $s$ all the way to $s=1$. Furthermore, all CFT data at $s=1$ are analytically computable from the free scalar description given above. Perturbative corrections to this data at small $\delta$ are constrained by the crossing equations~\eqref{eq:crossing} and the OPE relation~\eqref{eq:opeRelation}. We use this principle to uniquely fix the perturbative corrections to several orders.

Our only assumptions are: 1)~The CFT data agree with the free scalar model of the previous section at $\delta = 0$. 2)~They are described by asymptotic series in nonnegative powers of $\sqrt{\delta}$. 3) The two $\mathbb{Z}_2$-odd operators which become $\hat{\sigma}_3$ and $\im\partial\phi/\sqrt{2}$ at $\delta= 0$ behave like $\sigma$ and $\chi$ to all orders. That is, $\Delta_\sigma = \delta/2$, $\Delta_{\chi} = 1-\delta/2$, and the OPE relation~\eqref{eq:opeRelation} is valid.

From this starting point, we must first identify the two linear combinations $\cO_{-} = \cos\theta\,\cO_h -\sin\theta\,\cO_g$, $\cO_{+} = \sin\theta\,\cO_h +\cos\theta\,\cO_g$ which become $\Delta$-eigenstates for $\delta=0^{+}$. To that end, we consider the crossing equation~\eqref{eq:crossing} for $\langle\sigma\sigma\cO_{\pm}\cO_{\pm}\rangle$. To $O(\sqrt{\delta})$, only $\mathds{1}$ and $\cO_\pm$ contribute on the LHS while only $\sigma$ and $\chi$ contribute on the RHS \footnote{Note that at $\delta = 0$, we have the \emph{exact} OPE $\sigma(x)\chi(y) =\cO_{h}(y)= \cos\theta\,\cO_{-}(y) + \sin\theta\,\cO_{+}(y)$.}.  \eqref{eq:crossing} then implies $\theta=\pi/4$ and $\Delta_{\pm} = 1\pm \sqrt{2\delta}+O(\delta)$, in perfect agreement with the RG calculation~\eqref{eq:dimOpm}! This match only works if $c_{hgg} = \sqrt{2}$ at $\delta=0$, which indeed follows from $\langle\partial\phi(x_1) e^{\im\phi(x_2)} e^{-\im\phi(x_3)}\rangle_0=-2\im/|x_{12}x_{13}x_{23}|$. We further learn from the same crossing equation that $c_{\sigma\chi\pm}=1/\sqrt{2} + O(\sqrt{\delta})$ and $c_{\sigma\sigma\pm} = \pm\sqrt{\delta/2} + O(\delta)$.

At higher orders in $\sqrt{\delta}$, the four-point correlators receive contributions from infinitely many conformal blocks. Fortunately, all scaling dimensions tend to integers as $\delta\rightarrow 0$, which allows us to solve the infinite-dimensional problem analytically. The key tool are the analytic functionals of~\cite{Mazac:2016qev}, particularly their close cousins, introduced in Section 2 of~\cite{Mazac:2019shk}. We used them to isolate the contributions of individual conformal blocks and thus solve for the CFT data order by order in $\sqrt{\delta}$, similarly to the perturbative bootstrap of~\cite{Mazac:2018ycv}.

The bootstrap then correctly reproduces the RG result~\eqref{eq:dimOpm} up to $O(\delta)$ and furthermore determines the OPE coefficients $c_{ijk}$ between operators $\sigma$, $\chi$, and $\cO_{\pm}$. In particular, we find the following prediction for $\langle\sigma\sigma\cO_{\pm}\rangle$
\be
c_{\sigma\sigma\pm} =  \pm \frac{\sqrt{\delta}}{\sqrt{2}} - \frac{15}{16}\delta \mp \left(\frac{\pi ^2}{3}-\frac{543}{128}\right)\frac{\delta^{3/2}}{2^{3/2}} + O(\delta^2)\,.
\label{eq:opeSSPM}
\ee
We were able to reproduce it using RG in our model up to $O(\delta)$. Additional conformal bootstrap predictions are included in Appendix B.

Notice that the CFT data exhibits a symmetry under $\sqrt{\delta}\rightarrow -\sqrt{\delta}$ and $\cO_{-}\leftrightarrow\cO_{+}$. In other words, as we circle in the complex $s$ plane around $s=1$, we come back to the same CFT but $\cO_{-}$ and $\cO_{+}$ switch places.

\paragraph{\bf Conclusions.}
In this letter, we have put forward a duality between the long-range quartic model~\eqref{eq:action-LRI} and the impurity model~\eqref{eq:Z}: we claim that they both capture the same 1d CFT via an interacting IR fixed point for $s\in(1/2,1)$. While for~\eqref{eq:action-LRI}, the fixed point is weakly coupled near $s=1/2$ and strongly coupled near $s=1$, the situation is reversed for~\eqref{eq:Z}. Directly at $s=1$, our model becomes an exactly solvable conformal boundary condition, thus allowing us to identify the complete spectrum of scaling dimensions and OPE coefficients of the critical 1d long-range Ising model in the $s\rightarrow 1$ limit.

We have corroborated our proposal by RG and bootstrap calculations. The latter only requires information from the solvable $s=1$ theory as a seed, therefore providing a very powerful and independent validation.

The proposed duality solves the previously open problem of finding  a weakly coupled description of the 1d LRI model near $s=1$. Building on the physical picture of Anderson, Yuval and Kosterlitz, based on kink-antikink pairs, our duality provides a genuine field theory valid for $s < 1$. It is amenable to both RG and bootstrap techniques, allowing the computation of a wealth of CFT data near the crossover.

The point $s=1$ demarcates the long-range to short-range crossover, in the sense that the LRI model has no phase transition for $s>1$ at positive temperature, as is the case for the 1d SRI model.
In $d>1$, a weakly coupled description of the corresponding crossover was found in~\cite{Behan:2017dwr,Behan:2017emf}. The extension to $d=1$ was unknown until now, because of the absence of a critical SRI CFT in 1d. An important difference with respect to $d>1$ is that in $d=1$, the temperature becomes a marginal variable at the crossover. Equivalently, the leading invariant relevant operator $\cO_-$ becomes marginal at $s=1$, where the phase transition is of BKT type~\cite{Kosterlitz_2016}.

Notice that our $\s$ and $\chi$ fields play the same role as the eponymous fields in \cite{Behan:2017dwr,Behan:2017emf}, where their $\s$ is the SRI field, and $\chi$ an auxiliary GFF;  we also recognize our $\cO_h$ as the analogue of their $\cO= \s\chi$. Besides this analogy, our situation is rather different, enriched particularly by the presence of vertex operators. Therefore, $\s$ and $\chi$ are constructed in a different way, and $\cO_g$ must be included in the action~\eqref{eq:Z}. We also stress that the restriction to the $\mathrm{U}(1)$-invariant sector of our model plays an important role in the duality, as such a symmetry does not appear in the LRI model. We will return to these points in~\cite{long}.

\

\paragraph{Acknowledgments.}
We thank C. Behan, A. Gimenez-Grau, M. Paulos, L. Rastelli, S. Rychkov, J. Vo\v{s}mera and Zechuan Zheng for discussions. EL is supported by the European Union (ERC, QFT.zip project, Grant Agreement no. 101040260). PvV is funded by the European Union (ERC, FUNBOOTS, project number 101043588). Views and opinions expressed are however those of the authors only and do not necessarily reflect those of the European Union or the European Research Council Executive Agency. Neither the European Union nor the granting authority can be held responsible for them.

\appendix

\section{Appendix A -- Nonperturbative CFT constraints}
\label{app:constraints}

Let us provide a brief review of the two types of nonperturbative constrained satisfied by the critical 1d LRI CFT data for any $s\in(1/2,1)$, namely the crossing equations and the OPE relation.

Suppose $\cO_{i,j,k,\ell}$ are parity-even primary operators in a 1d CFT. We consider their four-point correlator $\langle\cO_{i}(x_1)\cO_{j}(x_2)\cO_{k}(x_3)\cO_{\ell}(x_4)\rangle$ for $x_1>x_2>x_3>x_4$. Thanks to the $\mathrm{PSL}_2(\mathbb{R})$ symmetry, it can be written as
\be
\begin{aligned}
&\langle\cO_{i}(x_1)\cO_{j}(x_2)\cO_{k}(x_3)\cO_{\ell}(x_4)\rangle 
= x_{12}^{-\Delta_1-\Delta_2+\Delta_3+\Delta_4}\\
&\times
x_{13}^{-2\Delta_3}
x_{24}^{\Delta_1-\Delta_2-\Delta_3-\Delta_4}
x_{14}^{-\Delta_1+\Delta_2+\Delta_3-\Delta_4}
\cG_{ijk\ell}(z)\,,
\end{aligned}
\ee
where $x_{ij}=x_i-x_j$ and $z = \frac{x_{12}x_{34}}{x_{13}x_{24}}\in(0,1)$ is the cross-ratio. By applying the OPE alternatively to $\cO_{i}(x_1)\cO_{j}(x_2)$ or $\cO_{j}(x_2)\cO_{k}(x_3)$, we obtain the s-channel and t-channel expansions:
\be
\begin{aligned}
&\cG_{ijk\ell}(z)\\
&= \sum\limits_{m}c_{ijm}c_{k\ell m}(-1)^{J_m} G^{\Delta_i,\Delta_j,\Delta_k,\Delta_\ell}_{\Delta_m}(z)\\
&=
\sum\limits_{m}c_{jkm}c_{\ell i m}(-1)^{J_m} G^{\Delta_i,\Delta_\ell,\Delta_k,\Delta_j}_{\Delta_m}(1-z)\,.
\end{aligned}
\ee
Here
\be
\begin{aligned}
&G^{\Delta_1,\Delta_2,\Delta_3,\Delta_4}_{\Delta_5}(z) =
z^{\Delta_5-\Delta_3-\Delta_4}\\
&\times{}_2F_1(\Delta_5-\Delta_1+\Delta_2,\Delta_5+\Delta_3-\Delta_4;2\Delta_5;z)
\end{aligned}
\ee
are the conformal blocks and $(-1)^{J_{m}}$ is the parity of $\cO_m$.

To derive the OPE relations, recall that operators $\sigma$ and $\chi$ are related by the shadow transform~\cite{Simmons-Duffin:2012juh}
\be
\sigma(x) = A \int\limits\frac{\chi(y)}{|y-x|^{2\Delta_\sigma}}dy\,.
\label{eq:shadow}
\ee
It follows that OPE coefficients involving $\sigma$ and $\chi$ are related to each other. Indeed, let $\cO_{i,j,k,\ell}$ be primary operators of dimensions $\Delta_{i,j,k,\ell}$ and parities $(-1)^{J_i,J_j,J_k,J_\ell}$ and let $a_{ij} = \frac{1-(-1)^{J_i+J_j}}{2}$. It follows from~\eqref{eq:shadow} that \cite{Paulos:2015jfa,Behan:2018hfx}
\be
\begin{aligned}
&\frac{c_{\sigma ij}c_{\chi k\ell}}{c_{\chi ij}c_{\sigma k\ell}}\\
&=\frac{\Gamma \left(\frac{\Delta_\sigma+\Delta_i-\Delta_j+a_{ij}}{2}\right)\Gamma \left(\frac{\Delta_\sigma-\Delta_i+\Delta_j+a_{ij}}{2}\right)}{\Gamma \left(\frac{1-\Delta_\sigma+\Delta_i-\Delta_j+a_{ij}}{2}\right)\Gamma \left(\frac{1-\Delta_\sigma-\Delta_i+\Delta_j+a_{ij}}{2}\right)}\\
&\quad\times\frac{\Gamma \left(\frac{1-\Delta_\sigma+\Delta_k-\Delta_\ell+a_{k\ell}}{2}\right)\Gamma \left(\frac{1-\Delta_\sigma-\Delta_k+\Delta_\ell+a_{k\ell}}{2}\right) }{ \Gamma \left(\frac{\Delta_\sigma+\Delta_k-\Delta_\ell+a_{k\ell}}{2}\right)\Gamma \left(\frac{\Delta_\sigma-\Delta_k+\Delta_\ell+a_{k\ell}}{2}\right)}\,.
\end{aligned}
\label{eq:opeRelation}
\ee
Reference \cite{Behan:2023ile} provides an alternative derivation of these OPE relations from the analyticity requirement discussed in \cite{Lauria:2020emq}.

\section{Appendix B -- Additional bootstrap results}
\label{app:results}
By expanding the crossing equation of $\langle\sigma\sigma\cO_{\pm}\cO_{\pm}\rangle$ to $O(\delta)$, disentangling the contributions of different operators using analytic functionals, and finally applying the OPE relation~\eqref{eq:opeRelation}, we derived~\eqref{eq:dimOpm},~\eqref{eq:opeSSPM} as well as 
\be
\begin{aligned}\label{eq:OPEs}
c_{\sigma\chi\pm} &= \frac{1}{\sqrt{2}}\mp\frac{\sqrt{\delta}}{16}-\frac{\delta}{256 \sqrt{2}} + O(\delta^{3/2})\\
c_{\chi\chi\pm} &= -\frac{\pi ^2 \delta }{2}\mp \frac{11 \pi ^2 \delta ^{3/2}}{16 \sqrt{2}}+O(\delta^2)\,.
\end{aligned}
\ee
We verified the result for $c_{\sigma\chi\pm}$ using RG in the model~\eqref{eq:Z} up to $O(\sqrt{\delta})$. The same bootstrap method also produces the following predictions for $\langle\cO_{\pm}\cO_{\pm}\cO_{\pm}\rangle$
\be
\begin{aligned}
c_{---} &= +\frac{3}{2} - \frac{39}{16\sqrt{2}}\sqrt{\delta} + O(\delta)\\
c_{--+} &= -\frac{1}{2} - \frac{1}{16\sqrt{2}}\sqrt{\delta} + O(\delta)\\
c_{-++} &= - \frac{1}{2} + \frac{1}{16\sqrt{2}}\sqrt{\delta} + O(\delta)\\
c_{+++} &= +\frac{3}{2} + \frac{39}{16\sqrt{2}}\sqrt{\delta} + O(\delta)\,.
\end{aligned}
\ee
These also agree with RG up to $O(\sqrt{\delta})$.

Finally, we studied the four-point correlator $\langle\sigma\sigma\sigma\sigma\rangle$ as a function of the cross ratio $z$. We were able to bootstrap it up to $O(\delta^2)$:
\be
\begin{aligned}
&\cG_{\sigma\sigma\sigma\sigma}(z) = 1 - \log[z(1-z)]\delta\\
&+\left[
-6 \text{Li}_3\left(\tfrac{z}{z-1}\right)+4\log (\tfrac{1-z}{z}) \text{Li}_2(z)\right.\\
&\quad\, + \log ^3(1-z)+\frac{1}{2}\log ^2(z)+\frac{1}{2}\log ^2(1-z)\\
&\quad\,-\log ^2(z) \log (1-z)
+\frac{1}{3} \pi ^2 \log (1-z)\\
&\left.\quad\,+\frac{1}{2} \log (z)\log (1-z)
\right]\delta^2+O(\delta^3)\,.
\end{aligned}
\ee
Crossing symmetry $\cG_{\sigma\sigma\sigma\sigma}(z)=\cG_{\sigma\sigma\sigma\sigma}(1-z)$ is manifest up to $O(\delta)$ and follows from polylogarithm identities at $O(\delta^2)$.

%----- Bibliography ----------------------

\bibliographystyle{apsrev4-2}
\bibliography{Refs}

%---------------------------------------------
%---------------------------------------------

\end{document}